\documentclass[12pt]{article}
\usepackage[dvips]{graphicx}
\usepackage{amsmath,amsfonts,amssymb,amsthm,color}

\usepackage{natbib}

\newtheorem{theorem}{Theorem}
\newtheorem{lemma}{Lemma}

\newtheorem{remark}{Remark}
\newtheorem{proposition}{Proposition}

\newtheorem{definition}{Definition}
\newtheorem{example}{Example}

\usepackage{bm}
\bmdefine{\Bt}{t}
\bmdefine{\BX}{X}
\bmdefine{\BY}{Y}
\bmdefine{\BZ}{Z}
\bmdefine{\BB}{B}
\bmdefine{\BM}{M}
\bmdefine{\BD}{D}
\bmdefine{\Bi}{i}
\bmdefine{\Bj}{j}
\bmdefine{\Bx}{x}
\bmdefine{\By}{y}
\bmdefine{\Bz}{z}
\bmdefine{\Bv}{v}
\bmdefine{\Bw}{w}
\bmdefine{\Bn}{n}
\bmdefine{\Ba}{a}
\bmdefine{\Bb}{b}
\bmdefine{\Bc}{c}
\bmdefine{\Be}{e}
\bmdefine{\Bu}{u}
\bmdefine{\Bp}{p}
\bmdefine{\Bzero}{0}
\bmdefine{\Bone}{1}

\newcommand{\Z}{{\mathbb Z}}
\newcommand{\R}{{\mathbb R}}

\def\comment#1{\textit{[#1]}}
\def\comment#1{}

\usepackage[small,bf]{caption}
\begin{document}

\title{Parametric $k$-best alignment}
\author{Peter Huggins \\ 
        Lane Center for Computational Biology \\ 
        Carnegie Mellon University \\  
          \, \\
         Ruriko Yoshida \\
         Department of Statistics \\
            University of Kentucky
         }
\date{}
\maketitle

\abstract{

Optimal sequence alignments depend heavily on alignment scoring parameters.  Given input sequences, {\em parametric alignment} is the well-studied problem that asks for all possible optimal alignment summaries as parameters vary, as well as the {\em optimality region} of alignment scoring parameters which yield each optimal alignment.  
But biologically correct alignments might be {\em suboptimal} for all parameter choices.  Thus we extend parametric alignment to {\em parametric $k$-best alignment}, 
which asks for all possible $k$-tuples of $k$-best alignment summaries $(s_1, s_2, \ldots, s_k)$, as well as the {\em $k$-best optimality 
region} of scoring parameters which make $s_1, s_2, \ldots, s_k$ the top $k$ summaries.  By exploiting the integer-structure of alignment summaries, we show that, astonishingly, the complexity of parametric $k$-best alignment is only polynomial in $k$.  
Thus parametric $k$-best alignment is tractable, and can be applied at the whole-genome scale like parametric alignment.  

}
\vskip 0.2in

\noindent
Corresponding author: \\
name: Peter Huggins \\
email address: phuggins@andrew.cmu.edu \\
address: \\

\pagebreak

\section{Introduction}

In pairwise sequence alignment, we are given a pair of homologous sequences $\sigma_1, \sigma_2$, and each alignment $\mathcal{A}$ of $\sigma_1, \sigma_2$ 
is endowed with an {\em alignment summary} $s(\mathcal{A}) \in {\bf Z}^d$ that records various features of $\mathcal{A}$, such as the number of mismatches and the number of spaces.  Throughout we will assume the dimension $d$ of alignment summaries is fixed, and that sequences are of length $O(n)$. 
The {\em score} of alignment $\mathcal{A}$ is defined to be $c \cdot s(\mathcal{A})$, where $c$ is a fixed vector of alignment scoring parameters. A {\em (global) optimal alignment} is any alignment $\mathcal{A}$ which maximizes the alignment score.  For most choices of alignment summary model, the optimal alignment summary can be computed in $O(n^2)$ time by the Needleman--Wunsch (NW) algorithm \cite{NW}, once the value $c$ of the alignment scoring parameters is given.

The choice of $c$ reflects relative frequencies of indels and different types of point mutations during sequence evolution.  
The optimal alignment is heavily dependent on the choice of $c$, and yet in practice the ``biologically correct'' choice of $c$ is not known. 
 Given sequences $\sigma_1, \sigma_2$, the space of alignment scoring parameters partitions into {\em optimality regions}.  
Parameter values in the same optimality region give rise to the same optimal alignment summary.  {\em Parametric alignment} \cite{berndlior} is the problem of determining 
all possible optimal alignment summaries that arise as $c$ varies, and also computing the optimality region for each optimal summary.  

Parametric alignment is a well-studied subject (see  \cite{Fernandez-Baca2000,Fernandez-Baca2002,Fernandez-Baca2005,Gusfield1994,Gusfield1996,Waterman1992}), and surprisingly tractable \cite{berndlior}:  
Pachter and Sturmfels proved that there are $O(n^{\frac{d(d-1)}{d+1}})$ optimality regions.  
In \cite{Gusfield1996, plos}, oracle-based methods for computing optimality regions are presented, which repeatedly run the NW algorithm with different
choices of scoring parameters $c$ to find new optimal alignments.  Despite the nice $O(n^{\frac{d(d-1)}{d+1}})$ bound on the number of optimality regions, it has been speculated \cite{bacachapter} that the required number of NW calls might be as high as $\Theta(n^{\frac{d^2(d-1)}{2(d+1)}})$.
One purpose of this paper is to point out that, actually, 
\begin{theorem}\label{thm1}
Existing oracle-based methods for parametric alignment only use $O(n^{\frac{d(d-1)}{d+1}})$ calls to the NW algorithm.
\end{theorem}

Thus parametric alignment is much more tractable than previously thought.
Nevertheless, one major shortcoming of parametric alignment is that it ignores {\em nearly optimal} alignments that are never optimal for any choice of scoring parameters.  Nearly optimal alignments have been studied before (see \cite{naorsuboptimal93, vingron96} and references within), but not in a parametric setting.  Thus we propose {\em parametric $k$-best alignment}, which studies how the $k$-best alignment summaries vary with parameters.  We consider two variants of the problem:  

\begin{itemize}
\item {\em Ordered} parametric $k$-best alignment:  Compute the collection of all ordered subsets of $k$ distinct alignment summaries $( s_1, \ldots, s_k )$ which can become the $k$-best summaries $c \cdot s_1 > c \cdot s_2 > \ldots > c \cdot s_k > \ldots$ under some choice of $c$.  For each such subset $( s_1, \ldots, s_k )$, find all $c$ such that $c \cdot s_1 \geq \ldots \geq c \cdot s_k \geq \ldots$ are the $k$-best summaries.
\item {\em Unordered} parametric $k$-best alignment:  Same problem, but the ordering of the $k$-best summaries $\{ s_1, \ldots, s_k \}$ is ignored.  
\end{itemize}

The output of parametric $k$-best alignment is a decomposition of the space of 
alignment scoring parameters into {\em $k$-best optimality regions}.  All scoring parameters in a $k$-best optimality region yield the 
same list of $k$-best distinct alignment summaries.  

Although parametric $k$-best alignment is a natural extension of parametric alignment, there are two major difficulties which have prevented 
its study:

\begin{itemize}
\item The structure of $k$-best optimality regions needs to be understood in order to systematically compute them, and 
\item  naively, we might worry that the number of $k$-best optimality regions is exponential in $k$.  
\end{itemize}

We first address the second point.  Notice that the total number of subsets of $k$ alignment summaries grows exponentially in $k$.  Indeed, if alignment summaries were arbitrary real-valued points, then the 
number of $k$-best optimality regions could be exponential in $k$ when $k < d/2$.  But alignment summaries are {\em integer points} contained in a small volume, 
and using this fact we have a remarkable result:

\begin{theorem}\label{thm1_5}
For fixed $k$, the number of $k$-best optimality regions is  $O(n^{\frac{d(d-1)}{d+1}})$, which matches the best known bound for the $k = 1$ case.  Specifically, for general $k$, the number of $k$-best optimality regions is  $O((kn)^{\frac{d(d-1)}{d+1}})$ for unordered parametric $k$-best alignment, and $O((k^2n)^{\frac{d(d-1)}{d+1}})$ for ordered parametric $k$-best alignment.
\end{theorem}

\begin{remark}\label{rk1}
Since there might be $\Omega (n^d)$ alignment summaries, Theorem \ref{thm1_5} says that, remarkably, the number of $k$-best optimality regions 
is sublinear in the worst-case number of alignment summaries if $k = o(n^{\frac{1}{d-1}})$. 
\end{remark}

In order to leverage Theorem \ref{thm1_5} and obtain fast parametric $k$-best alignment, we need to find those very few $k$-best optimality regions, 
without considering all possible subsets of $k$ summaries.  For standard parametric alignment ($k = 1$), the collection of optimality regions can be efficiently represented and computed via an object called the {\em alignment polytope}.  Polytopes are standard geometric objects which generalize polygons to higher dimensions.  In \cite{plos}, polytope construction software was used to efficiently compute alignment polytopes and solve parametric alignment.  

For $k > 1$, it was not clear whether $k$-best optimality regions could be represented by a polytope as in the $k = 1$ case.  
At the heart of our paper is the following affirmative result:

\begin{theorem}\label{thm2}
The collection of $k$-best optimality regions can be represented by a polytope called a $k$-set polytope.  
\end{theorem}

Specifically we define {\em ordered $k$-set polytopes} and {\em unordered $k$-set polytopes}, respectively, for ordered and unordered parametric $k$-best alignment.  For $k = 1$ the $k$-set polytopes are precisely the alignment polytope.  Our $k$-set polytopes eludicate the structure of $k$-best optimality regions, and allow us to generalize existing polytope algorithms for parametric alignment to the $k$-best setting.  
For standard parametric alignment, the oracle-based incremental polytope construction algorithm in \cite{plos} repeatedly runs the NW algorithm as a subroutine, with different choices of scoring parameters, in order to find new optimal alignment summaries.  Here we generalize the oracle-based incremental polytope construction algorithm to solve parametric $k$-best alignment.  In our generalized incremental algorithm, the standard NW algorithm is replaced with a $k$-best version of NW that finds the $k$-best alignment summaries, instead of just the optimal summary.  (The running time of the $k$-best NW algorithm is only a factor of $k$ larger than the running time of standard NW).  Our main result is:

\begin{theorem}\label{thm3}
There is an oracle-based incremental polytope construction algorithm to solve parametric $k$-best alignment.
The algorithm solves unordered parametric $k$-best alignment by calling the $k$-best version of the NW algorithm a total of  $O((kn)^{\frac{d(d-1)}{d+1}})$ times.  For ordered parametric $k$-best alignment, the $k$-best NW algorithm is called $O((k^2n)^{\frac{d(d-1)}{d+1}})$ times.  Besides NW calls, the rest of the algorithm's running time is $O((kn)^{\frac{2d(d-1)}{d+1}}))$ for unordered parametric $k$-best alignment, and $O((k^2n)^{\frac{2d(d-1)}{d+1}}))$ for ordered.
\end{theorem}

Furthermore, for $d \leq 3$ and $k = O(n^{1/4})$ the total running time of our algorithm is optimal, i.e. the running time is the same as running the NW algorithm once for each $k$-best optimality region. 

The most important feature of our algorithm's running time is that the dependence on $k$ is polynomial instead of exponential.  For small $k$ the running time of parametric $k$-best alignment is comparable to the best known bounds for the running time of standard parametric alignment.  
In \cite{plos} a whole-genome parametric alignment of {\em Drosophila} is presented, demonstrating how practical parametric alignment can be in practice.  Thus we are confident that parametric $k$-best alignment can be performed at the whole-genome scale as well, for not-too-large $k$.
The incremental polytope construction software {\tt iB4e} \cite{iB4e} can be used right out of the box to compute the necessary $k$-set polytopes, once the $k$-best version of the NW algorithm is written.


\section{Background on polyhedral geometry}

We begin by reviewing basic definitions and facts in polyhedral geometry.

\begin{definition}\label{def1}
The {\bf convex hull} of a set of points $V = \{v_1, \ldots, v_n\} \subset {\R}^d$ is the set $conv(V) = \{\sum c_i v_i \,\, | \,\, \sum c_i = 1, \,\, c_i \geq 0 \,\,\, \forall i \}$.  If $\sum c_i = 1$, and all $c_i \geq 0$, we say $\sum c_i v_i$ is a {\bf convex combination} of $V$.  
\end{definition}

\begin{definition}\label{def2}
A {\bf polytope} is a convex hull of any finite non-empty $V \subset {\R}^d$.
\end{definition}

The {\em dimension} of a polytope $P \subset {\R^d}$ is the dimension of its relative interior as a manifold.  
To avoid confusion between $d$ and $\dim P$, $d$ is called the {\em ambient dimension} of $P$.

\begin{definition}\label{def3}
Given a polytope $P \subset {\R}^d$ and a vector $c \in {\R}^d$, the {\bf face} $F_c \subset P$ is the set 
$F_c = \{x^* \in P \, | \, c \cdot x^* = \max_{x \in P} c \cdot x \}$.  By convention, the empty set is also considered to be a face of $P$
\end{definition}

Intuitively faces are the bounding extremities of the polytope.  Any face of a polytope $P$ is again a polytope, whose faces are also faces of $P$.  For most choices of $c$, the face $F_c$ will be a single point, which is called a {\em vertex} of $P$.  The 1-dimensional faces are called {\em edges}, and $(\dim P - 1)$-dimensional faces are called {\em facets}. 


\begin{definition}\label{def4}
Given a polytope $P \subset {\R}^d$ and a face $F \subset P$, the {\bf normal cone} $N(F)$ is the set of all vectors
$c$ for which $F_c \supseteq F$.
\end{definition}

In other words, the normal cone $N(F)$ is the set of all vectors $c$ such that $F$ weakly maximizes $c \cdot x$ over $P$.
There is a natural duality between faces and normal cones: for any two faces $F,G \subset P$ we have $G \subset F$ if and only if $N(G) \supset N(F)$.
The relative interiors of the normal cones of a polytope partition ${\R}^d$, and the collection of normal cones of all faces of $P$ is called the 
{\em normal fan} of $P$.

In this paper we will be interested in computing normal cones of vertices of a polytope $P$.  By duality, it suffices to know the facets of $P$ as we now explain.  For simplicity assume $\dim P = d$.  The facets $F_c$ of $P$ which contain $v$ give the set of vectors $c$ which generate the normal cone $N(v)$:  
\[
N(v) = \R_{\geq 0} \{ c \, | v \in F_c, \, F_c \hbox{ is a facet of } P \}
\]

For further reading on polytopes, see \cite{ziegler}.

\subsection*{Computing convex hulls}

Polytopes have been extensively studied in computational geometry, and many algorithms for convex hull construction have been devised \cite{Fukuda2001, qhull, edelsbrunnerbook}.
Unfortunately, traditional convex hull algorithms assume a point set $\mathcal{S}$ is explicitly given, for which $\hbox{conv}(\mathcal{S})$ is to be computed.  In sequence alignment we are presented with a quite different situation.  We don't know the set $\mathcal{S}$, but we seek to compute vertices and facets of $\hbox{conv}(\mathcal{S})$, and we have a fast {\em oracle} (e.g. the NW algorithm) which will find 
a vertex of $\hbox{conv}(\mathcal{S})$  that maximizes the dot-product with a given vector $c$.  The {\em incremental construction} algorithm and software reported in \cite{plos, iB4e} builds convex hulls efficiently in this setting.  Briefly put, the incremental construction algorithm repeatedly queries the vertex-finding oracle with different vectors $c$, adding one vertex at a time to the polytope, until all vertices of the convex hull are guaranteed to be found.  As shown in \cite{thesis}, we have

\begin{theorem}\label{thm4}
The incremental construction algorithm builds the convex hull of a point set $\mathcal{S}$, and all faces of $\hbox{conv}(\mathcal{S})$, given an oracle 
{\tt FindVertex}$(c)$ which maximizes given $c$ over $\mathcal{\mathcal{S}}$.  The oracle is queried $O(V+F)$ times, where $V$ and $F$ are the number of vertices and facets of $\hbox{conv}(\mathcal{S})$.  Besides oracle calls, the running time is $O(\ell_1 + \ldots + \ell_N)$, where $\ell_j$ is the number of faces of the convex hull after the first $j$ vertices are added.
\end{theorem}

\section{Ordered and unordered $k$-set polytopes}

It is straightforward to use polytopes as a tool for standard parametric alignment, simply by computing vertices and facets of the {\em alignment polytope}, i.e. the convex hull of alignment summaries \cite{berndlior}.  We now present special polytope constructions which are specifically designed for $k$-best alignment.  We begin with {\em ordered} parametric $k$-best alignment.

We will consider an (implicitly defined, but not explicity listed) set of $N$ alignment summaries $\mathcal{S} = \{ s_i \} \subset {\bf Z}^d$, and wish to compute the $k$-best summaries $(s_1, \ldots, s_k)$ in $\mathcal{S}$ with respect to a linear scoring scheme $c \cdot s_1 > c \cdot s_2 > \ldots $.  In particular we will be interested in computing all possibilities for the $k$-best summaries as $c$ varies (where the ordering of the summaries is taken into account).  We define polytopes $P_k$, which we call {\em ordered $k$-set polytopes}, whose vertices correspond to obtainable tuples of $k$-best summaries.

\begin{definition}
Given a set of $N$ alignment summaries $\mathcal{S} = \{ s_i \} \subset {\bf Z}^d$, let $(N)_k$ denote the set of all $N(N-1) \cdots (N-k+1)$ tuples of $k$ distinct indices $\sigma = (\sigma(1), \ldots, \sigma(k))$, where $\sigma(1), \ldots, \sigma(k) \in \{ 1, 2, \ldots, N \}$.  Notice that the ordering of the indices is taken into account.  The ordered $k$-set polytope $P_k$ for $\mathcal{S}$ is the convex hull 
\[
P_k = \hbox{conv} \{\sum_{i=1}^k (k+1-i) s_{\sigma(i)} \, | \sigma \in (N)_k \}.
\]
\end{definition}

\begin{center}
\begin{table}[ht]
\begin{center}
\begin{tabular}{llll}
$3 s_1 + 2 s_2 + s_3$ \,\, & $3 s_2 + 2 s_1 + s_3$ \,\, & $3 s_3 + 2 s_1 + s_2$ \,\, & $3 s_4 + 2 s_1 + s_2$ \\
$3 s_1 + 2 s_2 + s_4$ \,\, & $3 s_2 + 2 s_1 + s_4$ \,\, & $3 s_3 + 2 s_1 + s_4$ \,\, & $3 s_4 + 2 s_1 + s_3$ \\
$3 s_1 + 2 s_3 + s_2$ \,\, & $3 s_2 + 2 s_3 + s_1$ \,\, & $3 s_3 + 2 s_2 + s_1$ \,\, & $3 s_4 + 2 s_2 + s_1$ \\
$3 s_1 + 2 s_3 + s_4$ \,\, & $3 s_2 + 2 s_3 + s_4$ \,\, & $3 s_3 + 2 s_2 + s_4$ \,\, & $3 s_4 + 2 s_2 + s_3$ \\
$3 s_1 + 2 s_4 + s_2$ \,\, & $3 s_2 + 2 s_4 + s_1$ \,\, & $3 s_3 + 2 s_4 + s_1$ \,\, & $3 s_4 + 2 s_3 + s_1$ \\
$3 s_1 + 2 s_4 + s_3$ \,\, & $3 s_2 + 2 s_4 + s_3$ \,\, & $3 s_3 + 2 s_4 + s_2$ \,\, & $3 s_4 + 2 s_3 + s_2$ \\
\end{tabular}
\end{center}
\hskip 0.5in
\caption{Example of definition of ordered $3$-set polytope $P_3$, when $\mathcal{S}$ is a set of four points $s_1, s_2, s_3, s_4$.  In this case $P_3$ is the convex hull of $(4)_3 = 4 \cdot 3 \cdot 2 = 24$ points.  The 24 points are listed above.} \label{tb8}
\end{table}
\end{center}

The following theorem shows that the normal fan of $P_k$ gives precisely the ordered $k$-best optimality regions for the alignment summaries $\mathcal{S}$.  Thus computing vertices and facets of $P_k$ completely solves ordered parametric $k$-best alignment.

{
  \begin{theorem} \label{thm5}
   The normal cone of a point $\sum_{i = 1}^k (k+1-i) s_{\sigma(i)} \in P_k$ is the set of all $c$ satisfying 
   $c \cdot s_{\sigma(1)} \geq c \cdot s_{\sigma(2)} \geq \ldots \geq c \cdot s_{\sigma(k)}$, 
   and $c \cdot s_{\sigma(k)} \geq c \cdot s_j$ for all $j \notin \sigma$.  
  \end{theorem}
  }
 
\begin{proof}  See \cite{otherpaper}. \end{proof}

We now give analagous results for {\em unordered} parametric $k$-best alignment.

\begin{definition}
Given a set of $N$ points $\mathcal{S} = \{ s_i \} \subset {\bf R}^d$, let $\mathcal{S} \choose k$ denote the set of all $N \choose k$ subsets of $\mathcal{S}$ of size $k$.  The unordered $k$-set polytope $Q_k$ for $\mathcal{S}$ is
\[
Q_k = \hbox{conv} \{ \sum_{s \in A} s \, | \, A \in { \mathcal{S} \choose k } \}.
\]
\end{definition}

Unordered $k$-set polytopes have been previously studied \cite{edelshalving, fukudaksets}.  
We can modify Theorem \ref{thm5} to show that the normal fan of $Q_k$ gives the (unordered) $k$-best optimality regions for alignment summaries:

\begin{theorem}

 The normal cone of a point $\sum_{s \in A} s \in Q_k$ is the set of vectors $c$ which satisfy 
   $c \cdot s \geq c \cdot s'$ for all $s \in A$ and $s' \notin A$.

\end{theorem}

\begin{figure}[ht]

\begin{center}
\includegraphics[scale=0.6]{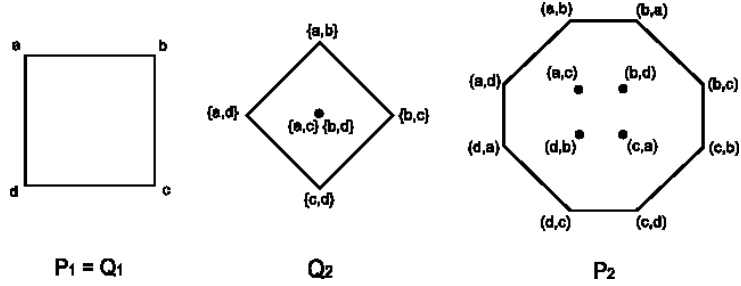} \\
\end{center}
\caption{Examples of unordered $k$-set polytopes $Q_k$ and ordered $k$-set polytopes $P_k$, for $\mathcal{S} = \{ a,b,c,d \} = $ vertices of a square.
         Points are labeled by the ordered/unordered $k$-set they represent.  Notice for example the point $2b + d$ is in the interior of $P_2$; this means that it is impossible for a linear scoring scheme to make $(b,d)$ the ordered top-2 points.  } \label{fig:kset_korder}
\end{figure}

\begin{example}\label{eg1}
Suppose $\mathcal{S}$ is the set of four vertices of a square.  Figure \ref{fig:kset_korder} shows some unordered $k$-set polytopes $Q_k$ and ordered $k$-set polytopes $P_k$ for $\mathcal{S}$.  
\end{example}

In parametric $k$-best alignment, the set of points $\mathcal{S}$ are alignment summaries, which 
are {\em integer points}.  In this case we can obtain remarkable  bounds on
the complexity of $k$-set polytopes.  

\begin{theorem}\label{thm7}
Suppose $\mathcal{S} \subset \Z^d$ is a set of $N$ integer points, and $k < N$.  Let $\mathbb{V}$ be the volume of $\hbox{conv}(\mathcal{S})$, and assume $\mathbb{V} > 0$.  If $V$ is any subset of vertices of the unordered $k$-set polytope $Q_k$ for $\mathcal{S}$, then the total number of faces of $\hbox{conv}(V)$ is $O((k^d \mathbb{V})^{(d-1)/(d+1)})$.  Similarly, if $W$ is any subset of the vertices of ordered $k$-set polytope for $\mathcal{S}$, then the total number of faces of the $\hbox{conv}(W)$ is $O((k^{2d} \mathbb{V})^{(d-1)/(d+1)})$.
\end{theorem}
\begin{proof}  See Appendix. \end{proof}

This concludes our treatment of $k$-set polytopes.  The rest of the paper gives applications to parametric $k$-best alignment, along with 
details on computation and implementation.

\section{Parametric $k$-best alignment}

We now aggregate the results of the previous sections to efficiently solve parametric $k$-best alignment.  For 
given sequences $\sigma_1, \sigma_2$ of length $O(n)$, let $\mathcal{S} = \{ s_1, \ldots, s_N \}$ be the set of all alignment summaries.  Each entry in an alignment summary counts a feature in the alignment, such as the number of occurences of a type of mismatch.  Thus $\hbox{conv}(\mathcal{S})$ has volume $O(n^d)$.

We first explain how to solve ordered parametric $k$-best alignment.  For $k \leq N$ let $P_k$ be the ordered $k$-set polytope for $\mathcal{S}$.
Given values for alignment scoring parameters $c$, the NW algorithm finds the optimal alignment summary by dynamic programming, keeping track of the optimal summary at each node in the alignment graph \cite{ASCB}.  Similarly, the NW algorithm can compute the top $k$ distinct alignment summaries  using the same type of dynamic programming recursion, keeping track of the top $k$ distinct summaries at each 
node in the alignment graph.  Thus we can define an oracle which, given $c$, will find the vertex $s^* \in P_k$ which maximizes $c \cdot s$ over $P_k$:

\begin{itemize}
\item Call the NW algorithm with scoring parameters $c$, to compute the top $k$ distinct alignment summaries $s_1, \ldots, s_k$ such that $c \cdot s_1 >  \ldots > c \cdot s_k > \ldots$.
\item Return $s^* := k s_1 + (k-1)s_2 + \cdots + s_k$.
\end{itemize}

Then Theorem \ref{thm4} says that the incremental polytope construction method will compute $P_k$ and its normal fan, calling the above oracle $O(V+F)$ times where $V$ and $F$ are the number of vertices and facets of $P_k$.  Since $\hbox{conv}(\mathcal{S})$ has volume $O(n^d)$, Theorem \ref{thm7} says that $V$ and $F$ are both $O((k^2n)^{\frac{d(d-1)}{d+1}})$.   Theorems \ref{thm4} and \ref{thm7} also tell us that besides oracle calls, the incremental polytope construction method takes no more than $O(V (k^2n)^{\frac{d(d-1)}{d+1}})$ time, which is $O( (k^2n)^{\frac{2d(d-1)}{d+1}})$.  Putting it all together, we have

\begin{theorem} \label{thm9}
Given sequences of length $O(n)$, let $V,F$ be the number of vertices and facets of the ordered $k$-set polytope for alignment summaries.  The incremental polytope construction method will solve ordered parametric $k$-best alignment in 
$O( (V+F) W(n,k) + (k^2n)^{\frac{2d(d-1)}{d+1}}))$ time, where $W(n,k)$ is the time required to run $k$-best NW once on sequences of length $O(n)$.
\end{theorem}
Typically $W(n,k) = \Theta (kn^2)$.  Table \ref{tb9} gives complexity bounds in this case for specific small values of $d$.

\begin{remark}\label{rk3}
The {\em upper bound theorem} for polytopes \cite{ziegler} says that the number of faces of a $d$-dimensional polytope is linear in the number of vertices if $d \leq 3$.  So if $d \leq 3$, and $W(n,k) = \Theta (kn^2)$, the running time of our algorithm is $O( Vkn^2 + V^2)$, and since $V = O(k^3 n^{3/2})$, the running time is thus $O(V \cdot W(n,k))$ when $k = O(n^{1/4})$.  {\em This is the same running time as the time required to run the NW algorithm once for each top $k$ ranking}.  Thus for $d \leq 3$ and $k = O (n^{1/4})$, our algorithm is an optimal oracle-based method.
\end{remark} 

\begin{remark}
When $d = 2$, at most one facet (edge) of the ordered $k$-set polytope is deleted when a new vertex is added (otherwise at least one vertex $v$ would also be deleted, contradicting that $v$ is a vertex).  Thus for $d = 2$ and $W(n,k) = \Theta (kn^2)$ the running time is  always the optimal $O(V \cdot W(n,k)) = O(k^{7/3} n^{8/3})$ for any $k$.
\end{remark}

\begin{remark}\label{rk2}
 In practice, we have observed that relatively few faces are created or destroyed when each new vertex of $P_k$ is found.  If an amortized $O(1)$ faces are created or destroyed when each new vertex of $P_k$ is found, then the incremental polytope construction method solves ordered parametric $k$-best alignment in $O( (V+F) \cdot W(n,k)) $ time, which is $O((k^2n)^{\frac{d(d-1)}{d+1}} \cdot W(n,k))$.  It is an important open question to determine the worst-case number of faces that can be created or destroyed during the incremental construction of $P_k$.
\end{remark}

We now explain how to solve unordered parametric $k$-best alignment.  The solution is analagous to ordered $k$-best alignment.  Let $Q_k$ be the unordered $k$-set polytope for $\mathcal{S}$.  We define an oracle which, given $c$, will find the vertex $s^* \in Q_k$ which maximizes $c \cdot s$ over $Q_k$:

\begin{itemize}
\item Call the NW algorithm with alignment scoring parameters $c$, to compute the top $k$ distinct alignment summaries $s_1, \ldots, s_k$ such that $c \cdot s_1 >  \ldots > c \cdot s_k > \ldots$.
\item Return $s^* := s_1 + s_2 + \cdots + s_k$.
\end{itemize}
Then, endowed with the above vertex-finding oracle, we have 

\begin{theorem} \label{thm10}
For sequences of length $O(n)$, let $V,F$ be the number of vertices and facets of the unordered $k$-set polytope for alignment summaries.  The incremental polytope construction method will solve unordered parametric $k$-best alignment in 
$O( (V+F) W(n,k) + (kn)^{\frac{2d(d-1)}{d+1}}))$ time.
\end{theorem}
Table \ref{tb9} gives specific bounds for small values of $d$, assuming $W(n,k) = \Theta (kn^2)$.

\begin{center}
\begin{table}[ht]
\begin{center}
\begin{tabular}{l|l|l|l|l}
$d$ & Output size             & Running time            & Output size             & Running time             \\ 
    & (unordered)             & (unordered)             & (ordered)               & (ordered)               
\\ \hline
& & & & \\   
2   & $O(k^{2/3}n^{2/3})$     & $O(k^{5/3}n^{8/3})$     & $O(k^{4/3}n^{2/3})$     & $O(k^{7/3}n^{8/3})$      \\ 
3   & $O(k^{3/2}n^{3/2})$     & $O(k^{5/2}n^{7/2} + k^3 n^3)$     & $O(k^{3}n^{3/2})$       & $O(k^{4}n^{7/2} + k^6 n^3)$        \\ 
4   & $O(k^{12/5}n^{12/5})$   & $O(k^{24/5}n^{24/5})$  &  $O(k^{24/5}n^{12/5})$  & $O(k^{48/5}n^{24/5})$    \\  

\end{tabular}
\end{center}
\hskip 0.5in
\caption{Running time and output complexity of ordered/unordered parametric $k$-best alignment for small dimensions, assuming the $k$-best version of the NW algorithm runs in $\Theta (kn^2)$ time.} \label{tb9}
\end{table}
\end{center}

\begin{remark}\label{rk4}
Analagous to Remark \ref{rk2}, the running time would be $O( (V+F) \cdot W(n,k))$ if we could prove that an amortized $O(1)$ faces are created or destroyed when each new vertex of $Q_k$ is found.
\end{remark}

\begin{remark}\label{rk5}
Analagous to Remark \ref{rk3}, if $d \leq 3$ the running time is  the optimal $O(V \cdot W(n,k))$ when $k = O(n)$ and $W(n,k) = \Theta (kn^2)$.  
\end{remark}

The software {\tt iB4e} reported in \cite{iB4e} can be used right out of the box to solve ordered and unordered parametric $k$-best alignment this way, once the oracles for finding vertices of $P_k$ and $Q_k$ are written.  A beta version of {\tt iB4e} was used in \cite{plos} to 
perform high throughput parametric alignment, and greatly outperformed the ``polytope semiring'' method reported in \cite{berndlior}.


\section{Discussion}

Although parametric alignment is a major improvement upon standard sequence alignment, parametric alignment ignores nearly optimal alignments.  Here we have extended parametric alignment to the $k$-best setting, determining how the top $k$ alignment summaries vary with  scoring parameters.  This allows for much more realistic parametric analysis of biological sequences.

Parametric alignment has enjoyed remarkably good complexity results, enabling whole-genome parametric analysis 
of {\em Drosophila} genomes \cite{plos}.  By extending the good complexity results to parametric $k$-best alignment, we believe parametric $k$-best alignment can be performed at the whole-genome scale as well.  As in \cite{plos}, such genome-scale parametric analysis will require standard preprocessing techniques that break up pairwise genomes into smaller reliably homologous subsequences.
  
In some applications, estimates of scoring parameters might be known along with confidence intervals on the estimates.  In this case 
we can restrict attention to optimality regions which intersect the confidence region for parameters.
It is possible to augment the vertex-finding oracle in {\tt iB4e} so that only optimal alignments whose optimality regions intersect a prescribed cone $C$ are found; other optimal alignments are completely avoided.  Details can be found in \cite{thesis}.  For example optimality regions could be restricted to a cone over a bounding box, as in \cite{Gusfield1996}.  Restricting the parameter space has the additional benefit of speeding up parametric $k$-best alignment, by reducing the number of optimality regions.  

The dimension $d$ of alignment summaries is the most prohibitive factor in the complexity of both parametric alignment and $k$-best alignment.  
But the curse of dimension is not nearly as bad as was speculated in \cite{bacachapter}.  
While some polytopes with $V$ vertices might have $\Theta(V^{\lfloor{d/2}\rfloor})$ faces, we have shown that $k$-set polytopes are special, and that the remarkable bounds on their number of vertices also applies to faces of all dimensions.  Thus parametric alignment and $k$-best alignment are much more tractable than previously thought.  This agrees with empirical observations, e.g. in \cite{plos} parametric alignment was demonstrated to be computationally practical for $d \leq 5$ {\em at the whole genome level}.  Our complexity results indicate that parametric $k$-best alignment will be similarly tractable.  Based on compututational experience with parametric alignment, we believe parametric $k$-best alignment will even be tractable for $d = 6,7$ when sequences are short.

It is important to note that restricting alignment summaries to have dimension $\leq 7$ prohibits the most general models of alignment scoring parameters.  For protein sequences, all but the most basic scoring matrices will yield $d > 7$.  Thus parametric alignment is not well-suited for protein sequence analysis.  Fortunately, for DNA sequences, popular scoring models such as 
those based on Jukes--Cantor, Kimura-2, and Kimura-3 scoring matrices will result in $d \leq 6$.  

Parametric alignment belongs to a more general class of algorithms called {\em parametric inference} algorithms for graph-based models \cite{berndlior}.  We note that the framework we have laid out here, extending parametric alignment to the $k$-best setting, can be adapted to perform parametric $k$-best inference in other graph-based models as well.  The remarkable complexity results we have proved can be extended to 
the parametric $k$-best inference setting as well.  Similarly the software {\tt iB4e} can be used to perform efficient parametric $k$-best inference, provided an oracle which performs $k$-best inference given scoring parameters.  Two important graph-based models in biology which can benefit from parametric $k$-best inference are hidden Markov models over discrete state spaces, and tree-models for single nucleotide evolution.  The vertex-finding oracles provided to {\tt iB4e} for these graph-based models would be the $k$-best Viterbi algorithm and $k$-best Felsenstein pruning algorithm respectively.

\section*{Acknowledgements}
R. Yoshida is supported by NIH R01 grant 1R01GM086888-01.

\pagebreak
\bibliographystyle{plain}
\bibliography{recom}

\pagebreak

\appendix

\section{Appendix}

Ordered and unordered $k$-set polytopes have rich structure which has still not been fully explored.  Here we recall a result from \cite{otherpaper} which shows that $k$-set polytopes can be intractable for general sets of points:  

\begin{proposition}\label{prop2}
Given a set of $N$ points $\mathcal{S} \subset \R^d$, let $f_k$ be the number of $k$-dimensional faces of $\hbox{conv}(\mathcal{S})$.  Then the ordered $k$-set polytope for $V$ has at least $(k+1)! f_k$ vertices.  Thus $k$-set polytopes have $\Theta(N^k)$ vertices in the worst case, if $k < d/2$.
\end{proposition}

However, alignment summaries are integer points.  Suppose $\mathcal{S} \subset \Z^d$ is a finite set of integer points.  Then as recalled in \cite{barany}, we have:

\begin{theorem}[Andrews--Barany]\label{thm6}
Let $\mathbb{V}$ be the volume of $\hbox{conv}(S)$.  If $\mathbb{V} > 0$, then the number of $k$-dimensional faces of $\hbox{conv}(S)$ is $O(\mathbb{V}^{(d-1)/(d+1)})$ for every $k$.
\end{theorem}

No such result is possible when $\mathcal{S}$ is an arbitrary set of real-valued points.

\subsection{Proof of Theorem \ref{thm7}}

The proof requires a lemma, proved in \cite{otherpaper}:

\begin{lemma}\label{lemma1}
Suppose $\mathcal{S} \subset \R^d$ is a set of $N \geq 1$ points.  If $1 \leq k < N$, then the $k$-set polytopes $Q_k$ and $P_k$ for $\mathcal{S}$ have the same dimension as $\hbox{conv}(\mathcal{S})$.
\end{lemma}

\begin{proof}[Proof of Theorem \ref{thm7}]   By definition of Minkowski sum, every point in the $k$-set polytope $Q_k$ is also a point 
in the $k$-fold Minkowsi sum $\hbox{conv}(S)^{\odot k}$, which equals the 
$k$-fold dilation $k \cdot \hbox{conv}(S) = \{ kx \, | \, x \in \hbox{conv}(S) \}$.  Thus the volume of $Q_k$ is no more than the volume of  $k \cdot \hbox{conv}(S)$, which is
$k^d{\mathbb V}$.  By the lemma, the volume of $Q_k$ is positive, so we can apply Theorem \ref{thm6}
and obtain that the number of faces of $Q_k$ is $O((k^d{\mathbb V})^{(d-1)/(d+1)})$.

Now, for the ordered $k$-set polytope $P_k$, we recall that $P_k = Q_1 \odot \ldots \odot Q_k$.  Since each $Q_j$ is a subset $j \cdot \hbox{conv}(S)$, we have that $P_k \subset k^2 \cdot \hbox{conv}(S)$.  Now $P_k$ is a lattice polytope of positive volume 
$\leq k^{2d}{\mathbb V}$.  So Theorem \ref{thm7} tells us that the number of faces of $P_k$ is $O((k^{2d}{\mathbb V})^{(d-1)/(d+1)})$.
\end{proof}

\end{document}